\pdfoutput=1
\documentclass[12pt,preprint]{aastex} 
\usepackage{natbib}  
\usepackage{subfig}
\shortauthors{Basri et al.}   
\shorttitle{Periodicity in \textit{Kepler} Target Stars} 
\begin{document} 

\title{Photometric Variability in \textit{Kepler} Target Stars. II. An Overview of Amplitude, Periodicity, and Rotation in First Quarter Data.} 

\author{ 
Gibor Basri\altaffilmark{1}, Lucianne M. Walkowicz\altaffilmark{1}, Natalie Batalha\altaffilmark{2}, Ronald L. Gilliland\altaffilmark{3}, Jon Jenkins\altaffilmark{2}, William J. Borucki\altaffilmark{2}, David Koch\altaffilmark{2}, Doug Caldwell\altaffilmark{2}, Andrea K. Dupree\altaffilmark{4}, David W. Latham\altaffilmark{4},Geoffrey W. Marcy \altaffilmark{1}, Soeren Meibom\altaffilmark{4}, Tim Brown \altaffilmark{5} 
} 

\altaffiltext{1}{Astronomy Department, University of California,  Berkeley, CA 94720} 
\altaffiltext{2}{NASA Ames Research Center, Moffett Field, CA 94035} 
\altaffiltext{3}{Space Telescope Science Institute, Baltimore, MD 21218} 
\altaffiltext{4}{Harvard-Smithsonian Center for Astrophysics, Cambridge, MA 02138} 
\altaffiltext{5}{Las Cumbres Observatory Global Telescope, Goleta, CA 93117} 

\begin{abstract} 
We provide an overview of stellar variability in the first quarter of data from the \textit{Kepler} mission. The intent of this paper is to examine the entire sample of over 150,000 target stars for periodic behavior in their lightcurves, and relate this to stellar characteristics. These data constitute an unprecedented study of stellar variability given its great precision and complete time coverage (with a half hour cadence). Because the full Kepler pipeline is not currently suitable for a study of stellar variability of this sort, we describe our procedures for treating the ``raw" pipeline data. About half of the total sample exhibits convincing periodic variability up to two weeks, with amplitudes ranging from differential intensity changes less than 10$^{-4}$ up to more than 10 percent. K and M dwarfs have a greater fraction of period behavior than G dwarfs. The giants in the sample have distinctive quasi-periodic behavior, but are not periodic in the way we define it. Not all periodicities are due to rotation, and the most significant period is not necessarily the rotation period. We discuss properties of the lightcurves, and in particular look at a sample of very clearly periodic G dwarfs. It is clear that a large number of them do vary because of rotation and starspots, but it will take further analysis to fully exploit this.
\end{abstract} 


\section{Introduction} 

The \textit{Kepler} mission provides an unprecedented set of data on the photometric variability of stars \cite{BoruckiScience}. It is unparalleled in its combination of photometric precision and time coverage. It observes a large sample of stars, which consist primarily of main sequence stars in the solar neighborhood. Although the primary mission is to discover exoplanets, clearly the stellar lightcurves are themselves a rich source of astrophysical information. Stars can vary their luminosity for a variety of reasons. Our orientation in this paper is to concentrate on stellar magnetic activity (primarily seen through starspot modulation). Questions of interest in this area include the stellar rotation periods, the differential rotation as a function of latitude, the  distribution of spot areal coverages, the distributions in longitude and latitude on different stars, the presence and distribution of active longitudes, the timescale for evolution of spots of different sizes, spot contrasts, and the behavior of all these as a function of stellar mass, age, and rotation. Of particular interest to the mission is the translation of photometric periods into stellar ages for planet-bearing stars. This paper is not intended to provide answers to any of the above questions, but rather to present the characteristics of the \textit{Kepler} dataset and its potential for making progress on these questions. Starspots have long been studied with either photometry or spectroscopy \cite{Strassmeier}. Ground-based coverage, however, tends to have many gaps, and the spot areal coverage has to be many times solar for a signal to be robust.  The promise of solving these deficiencies was well-illustrated first by the MOST \cite{MOST} and COROT \cite{COROT} missions, but with less precision and length of coverage than will be available over the \textit{Kepler} mission. We have discussed the general level of variability in the first quarter of \textit{Kepler} data in our first paper \cite{PaperI}; hereafter Paper I), where there is also a lengthier introduction to stellar variability in the context we are interested in here.  

\section{Observations and Data Reduction} 

The \textit{Kepler} Quarter 1 (Q1) observations took place over $\sim$33.5 days between May 13 and June 15 2009. This paper refers to data gathered in Q1, which was almost entirely released to the public on June 15, 2010. We use only the so-called Long Cadence data which is formed on-board the spacecraft by summing 270 6 second  integrations into successive 29.42 minute blocks (see \cite{Jenkins} for additional details). Q1 consists of 1639 such intervals from 13 May 2009 to 15 June 2009, which provide continuous coverage. 156,097 stars were observed, almost all of which belong to the sample of stars that are identified as the core exoplanet target list. The construction of this target list has been described by \cite{BatalhaTarg}. As described in the Release Notes, the data product which is relevant to this paper is the so-called raw data (identified by the keyword AP\_RAW\_FLUX in the FITS files). This is not really raw pixel data - it has been added up in each target aperture, corrected for background, flat fielding, cosmic ray removal, non-linearities, and there is some treatment of selected instrumental effects and data gaps. What remains in the raw photometry are other instrumental artifacts \cite{CaldwellInstr}, secular instrumental drifts \cite{Jenkins}, and intrinsic astrophysical variability.  

Subsequent steps in the \textit{Kepler} pipeline attempt to render the data more suitable for the primary mission: the detection of transits. Currently the methodology employed \cite{JenkinsPipe} is not suitable for  a study of stellar variability; one cannot be sure what components of variability have been removed, and the effects of reduction are different for different amounts of stellar variability. By comparing raw and corrected lightcurves, we determined that only raw lightcurves are suitable as a starting point for work that concentrates in detail on the variability due to stars (that may change with future versions of the pipeline). We elect to use a form of differential photometry, produced by dividing each lightcurve by its median value then subtracting unity. 

We next investigated whether the secular trends seen in almost all raw lightcurves have an instrumental component that can be easily distinguished from true stellar variability. We made a linear fit to each curve, and examined the slopes as a function of position within a given detector or across the focal plane. It is not surprising to find these slopes; the stars drift on the pixels due to pointing errors, positioning of the field, differential velocity aberration, and focal plane distortion due to thermal and other effects. The response of each pixel can vary. The psf of the \textit{Kepler} photometer has been intentionally designed to spread a star over several pixels, and apertures are intentionally larger than a target psf (both help with photometric precision). Stars at the edge of each target aperture can contribute more or less light over time depending on the direction of motion. Because of the great precision of \textit{Kepler} photometry, all these effects may be important, depending on the target star and neighboring star field being measured. Unfortunately, we conclude that to first order the overall effects on the measured intensity of a star are very local, and one cannot use the behavior of nearby stars to reliably remove trends in a given target. Of course, the linear trend in our fitting procedure could have a truly stellar component in some cases, and we have to acknowledge that in such cases it will be removed. 

After removal (by subtraction) of linear trends, it is clear that for many quiet stars, there remain curvatures in the lightcurve over the month of Q1  (there is no reason that instrumental secular effects should be purely linear). These also are not correlated spatially with each other, but it is clear that there is an instrumental component to them, because it is very common for such lightcurves with a linear fit removed to be high in the middle and low at the ends. With a quadratic fit removed, there was also some (though not as much) consistency in the slope of the curves at the ends. In most cases, therefore, we subtracted a third order polynomial from the lightcurve. This resulted in flat lightcurves for many of the quiet stars; again with the danger that in some cases an astrophysical signal with a timescale of a couple of weeks or longer has been removed. This means that we are likely missing some slower low-amplitude rotators.  

We also tested the effect of removing a third-order fit from pure sine curves of 1 to 4 cycles, constructed using the same cadence and timespan as the real data. For a single cycle (a period of 33.5 days), the third order fit crosses the sine wave 5 times and remains close to it. After subtraction one is left with a much smaller amplitude, and the periodogram has its maximum peak at about half the original period. The corrected lightcurve is much less affected for a 2-cycle curve (with an intrinsic period of 16.75 days). The amplitude after subtraction of a polynomial fit remains the same (with a slightly altered shape for the corrected lightcurve), and the inferred primary period is closer to 16 rather than 17 days. The effects for more cycles are even smaller. We conclude that our reduction procedure only allows valid discussion of primary periodicities of 16 days or less. Obviously, as Kepler continues to collect data this issue will be re-visited, along with development of methods for joining data across time intervals when the spacecraft has left and returned to the target field.  

A caveat to the removal of third order fits must be made in the case when a star is clearly quite variable primarily on a timescale of 10 days or more. That is to say, the corrections on quiet stars were never large in amplitude (not larger than about 10 times the rms fluctuations on short timescales), and essentially all of them were flattened with a third order polynomial. This indicates that the spacecraft does not generally produce signals with higher amplitude and higher frequency than the polynomial. We therefore made an exception to the removal of a third order fit in the following circumstance. After removal of the linear fit, the differential lightcurve can be analyzed for zero crossings (the number of times it goes from positive to negative or vice versa). For a quiet star with a flat lightcurve there will be many of these (albeit with small excursions from zero). For a star with substantial low frequency variability the number of zero crossings gives an indicator of what its timescale of variability is. We used this to avoid removing a third order polynomial in lightcurves first smoothed over 10 hours which then exhibit 10 or fewer zero crossings. This is clearly somewhat arbitrary, and could be refined in the future. It is worth noting that in the sine wave test described above, even the 4-cycle test would not have had a third-order polynomial removed, since it has too few zero crossings. Thus we believe that our procedure retains almost all of the variability we are primarily interested in.

Finally, there are a few cases (usually less than 5 per 1000) where the raw lightcurve contains a sudden large discontinuity. These are removed in the pipeline after the raw stage, but because we cannot use that version, we had to remove them ourselves. This was done by filtering all lightcurves with a routine that looked for large abrupt changes (a version of the routine we also use to look for eclipses and flares). This yielded few enough cases that they could be dealt with manually. It is still true, however, the best way to treat them  is not always obvious, and we employed three different strategies. In most cases, the discontinuity occurred somewhere in the middle of a not too variable lightcurve. In that case we removed a third order polynomial on each side of the discontinuity; these tended to match across the gap, but if they did not we forced a match. This procedure removes variability on a shorter timescale than one polynomial fit across the whole lightcurve.  

In other cases with more variable stars, it was clearly damaging to use the first procedure (the polynomials removed variations that are clearly stellar), so we simply removed the discontinuity by forcing a match across the gap, always adjusting the latter part of the curve to agree with the earlier part. This often resulted in a new residual linear slope, which we removed again. Finally, in some cases the discontinuity occurred near one end or the other, and we selected which of the above procedures gave a more plausible solution. Fortunately there were rather few discontinuities, although there remain some that are below our chosen detection threshold; these do not matter in the current analysis. This manual intervention consumes most of the reduction time; the rest of it (including making the measurements described in the next section) can be done in a few hours on a personal computer. 

Clearly all of these corrections have the potential to alter an astrophysical signal. For now we must accept that the photometry is not perfect (nor will it ever be), and look forward to the much longer measurement of each star that \textit{Kepler} will obtain, which should clarify in many cases what the right thing to do is. The purpose of this paper, in any case, is to produce a large statistical overview of the stellar photometry, rather than to model in detail particular stars. We are confident that the conclusions we draw do not depend too much on the details of the preliminary methodology we have employed in this paper. We are capturing most of the amplitude and characteristics of stellar variability on timescales less than about 2 weeks, and can distinguish between periodic and non-periodic behavior. We can see whether the observed behavior is repetitious over a month, and can detect flares and other short timescale phenomena.

\section{Analysis of Periodicity, Variability, and Rotation} 

Once the data were reduced as described in Section II, we proceeded to collect certain measurements on each lightcurve. Here we describe only those which proved useful to the aims of this paper. As in Paper I, one useful quantity is the variability range V$_{rng}$ (sometimes just called ``range"), which is intended to represent the basic level of photometric variability. We slightly changed the means of computing this quantity from Paper I, taking the unsmoothed differential lightcurve, sorting it by differential intensity, then computing the range between the 5$^{th}$ and 95$^{th}$ percentile of intensity (this tends to avoid including really anomalous excursions up or down). We express V$_{rng}$ in mmag for convenience (actually the units are differential intensity times 1085.84, which is nearly correct but not logarithmic like the actual magnitude scale). We determined the zero crossings in both the reduced lightcurve itself and when it is smoothed by 10 hours (20 points), tabulating both the number of crossings and the mean and median separation of crossings. For all segments between crossings (either above or below zero) we saved the mean and median of absolute maximum excursions and integrated fluxes. The high frequency noise HF$_{rms}$ in the lightcurve was calculated by subtracting a four-point boxcar smoothed version of itself from the lightcurve and computing the standard deviation of the result. 

We also computed a Lomb-Scargle periodogram for each lightcurve, using 400 points spanning a range of periods between one and 100 days. These parameters were chosen because the procedure uses a logarithmic period spacing that is denser at short periods, so we extend the period range beyond the data length to force sufficient sampling in the periods of interest. We do not make use of results for periods beyond half the data length (16 days) or for periods below 3 days (to concentrate on solar-type rotating stars). This is the most computationally intensive part of our analysis, and we rebinned the lightcurve to 2 hour bins to save time. It then only requires a few hours on a small computer to obtain periodograms for all 156,000 stars. We saved the position, height, and width of all the peaks in them. We collected the height and integrated strength of the highest two peaks, the shortest and longest significant periods, and the number of peaks at least 10\% as strong as the strongest one.  

\subsection{Samples of Periodic and Non-periodic Variables} 

\begin{figure} 
\centering
\includegraphics[width=0.75\textwidth]{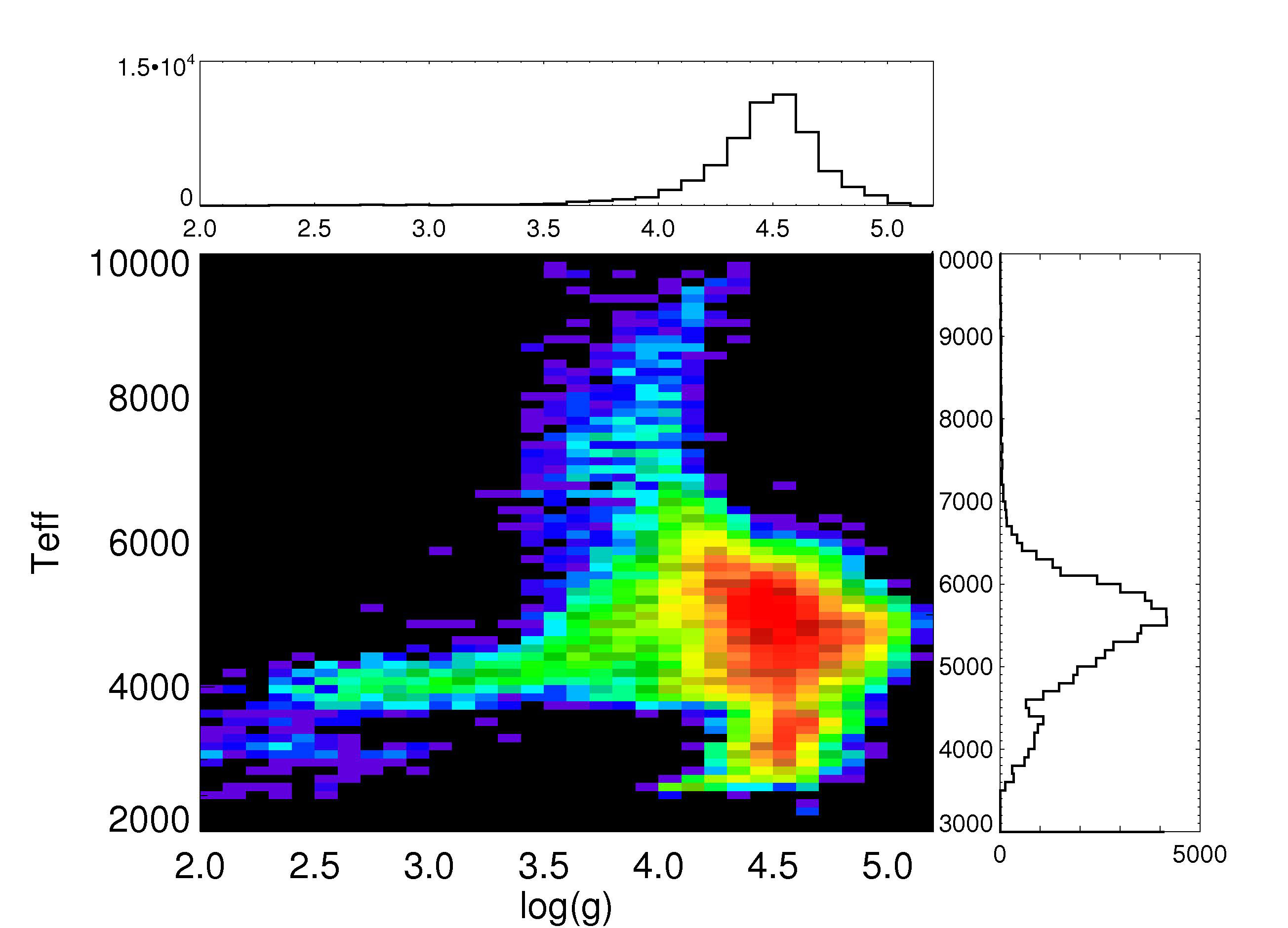}
\includegraphics[width=0.75\textwidth]{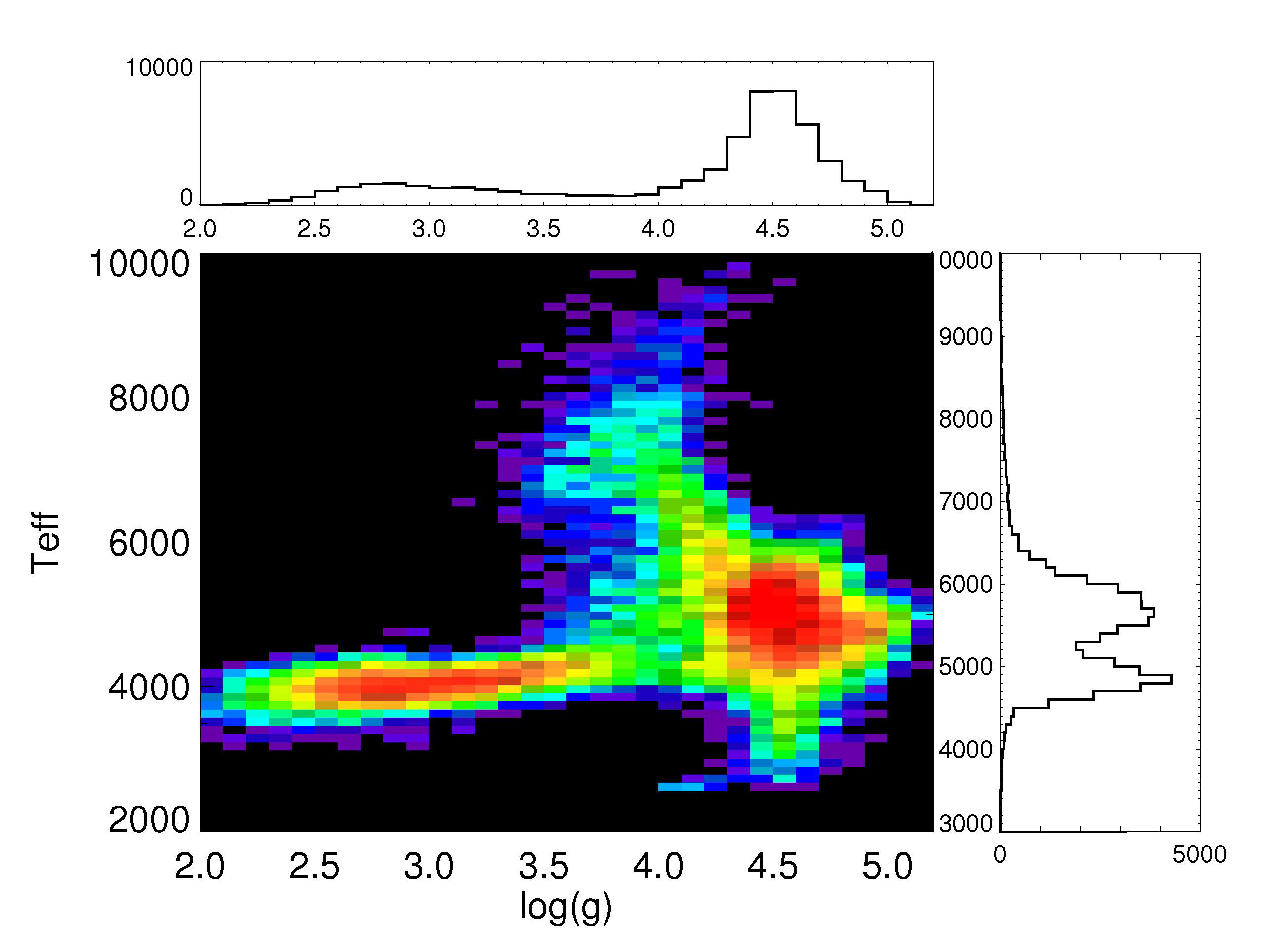}
\caption{Gravity vs. Temperature (using KIC parameters). The upper sample has stars with clear periodicity, while the lower sample is non-periodic. The periodic sample shows a stronger component of cool dwarfs. The non-periodic sample has post-main sequence stars (with low gravity); the hotter stars there tend to be pulsators with too high a frequency to trigger our filter for periods.  } 
\label{f1persample} 
\end{figure}

Our first division of the data was made on the basis of the strength of the strongest peak in the periodograms. After extensive manual examination, we found that strengths P$_{str}$ of 60 or more showed by eye a clear believable periodicity, those between 35 and 60 were marginal, and those below 35 did not appear to be period in an obvious way. The values of P$_{str}$ depend on the ratio of the periodic signal to the noise and on the number of  data samples and periods tested. The formal false alarm probability for our case is 1\% at P$_{str}$ of 13, and absurdly low for P$_{str}$ of 35, so we are being subjective but conservative in our judgment of what is a significant periodicity. The value of P$_{str}$ for a pure sine wave is around 800. In Figure \ref{f1persample} we show the Kepler Input Catalog (KIC) stellar parameters for the group of stars with P$_{str}$ above 60 and below 35 (we will refer to these as the periodic and non-periodic samples). There are features in the log(g)-T$_{eff}$ plane that are due to artifacts in the way the KIC was constructed and in target selection criteria for the exoplanet search \cite{BatalhaKIC}. We checked that the distribution of crowding factors (the fraction of light in target apertures that doesn't come from the intended target star) is very similar for the two samples; the median crowding factor is 0.135.   

There are some clear differences between the two samples. Giants are much more populous in the non-periodic sample. As described by \cite{BeddGiant} they tend to have complex variability; periodograms do not reach the level of P$_{str}$ we have set as our minimum threshold for significant periodicity (and exhibit multiple period peaks of comparable strength). By contrast, there are many solar-type stars in both the periodic and non-periodic samples. Cool dwarfs tend to show up preferentially as periodic. Hot stars also show up in both cases; part of the reason for this is that we have not tested for very short periods. Many of the hot pulsators are not flagged in our diagnostics but would be if we extended the search to shorter periods. There are more than 60,000 stars in our periodic sample, and nearly that many in the non-periodic sample. The marginal stars make up the roughly 35,000 remainder. The large number of stars with detected periods means that \textit{Kepler} will be quite powerful in measuring stellar rotation. We are only considering slightly over a month's data here and more than a year has already been collected. On one hand we should be able to detect spot rotations both longer in period and perhaps lower in amplitude as more data is available. On the other hand substantial issues remain to be resolved about how well stellar variability can be distinguished from instrumental effects, and how to patch across data segments where the stars are on different pixels. 

\begin{figure} 
\centering
\includegraphics[width=0.75\textwidth]{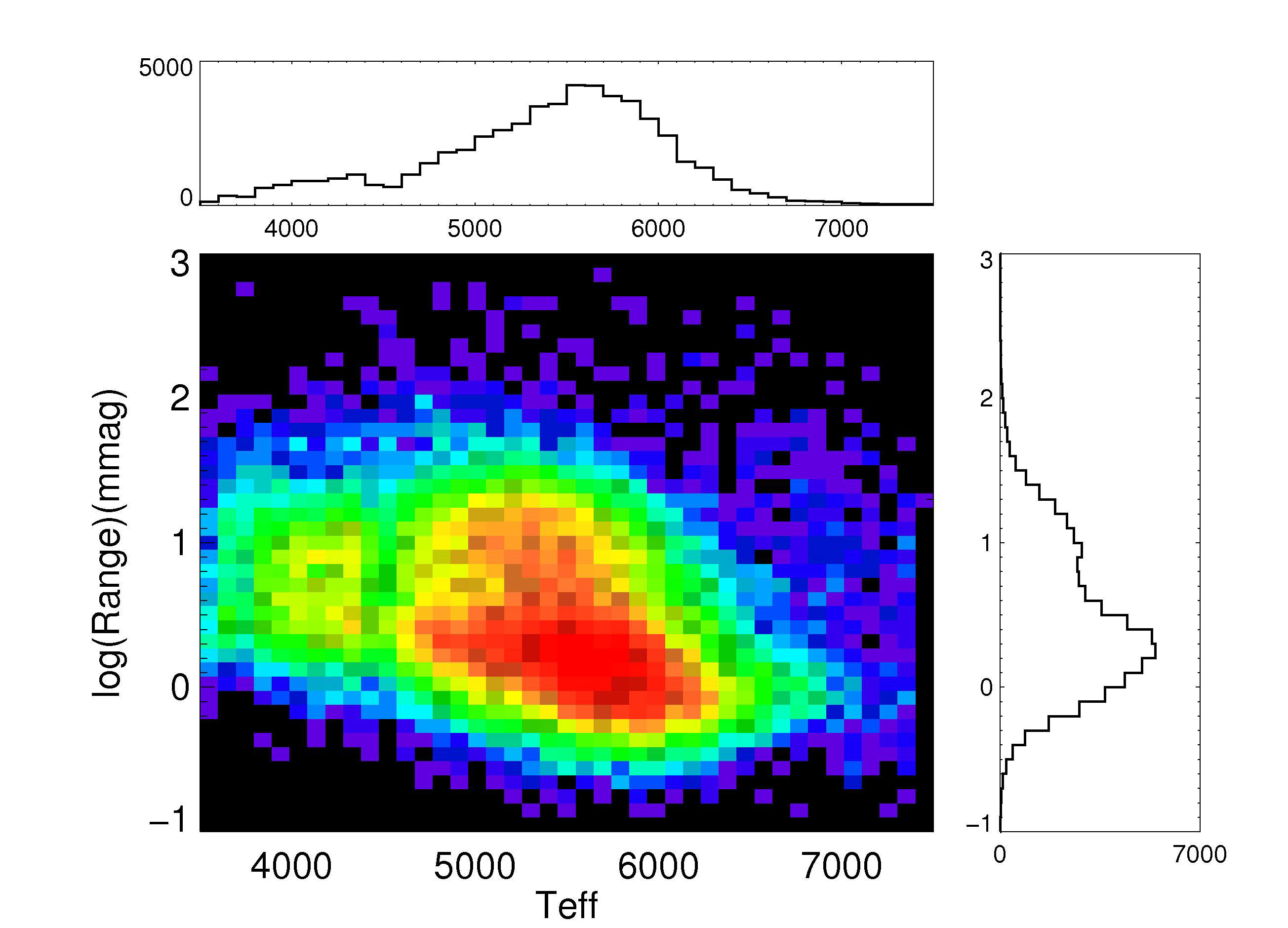} 
\includegraphics[width=0.75\textwidth]{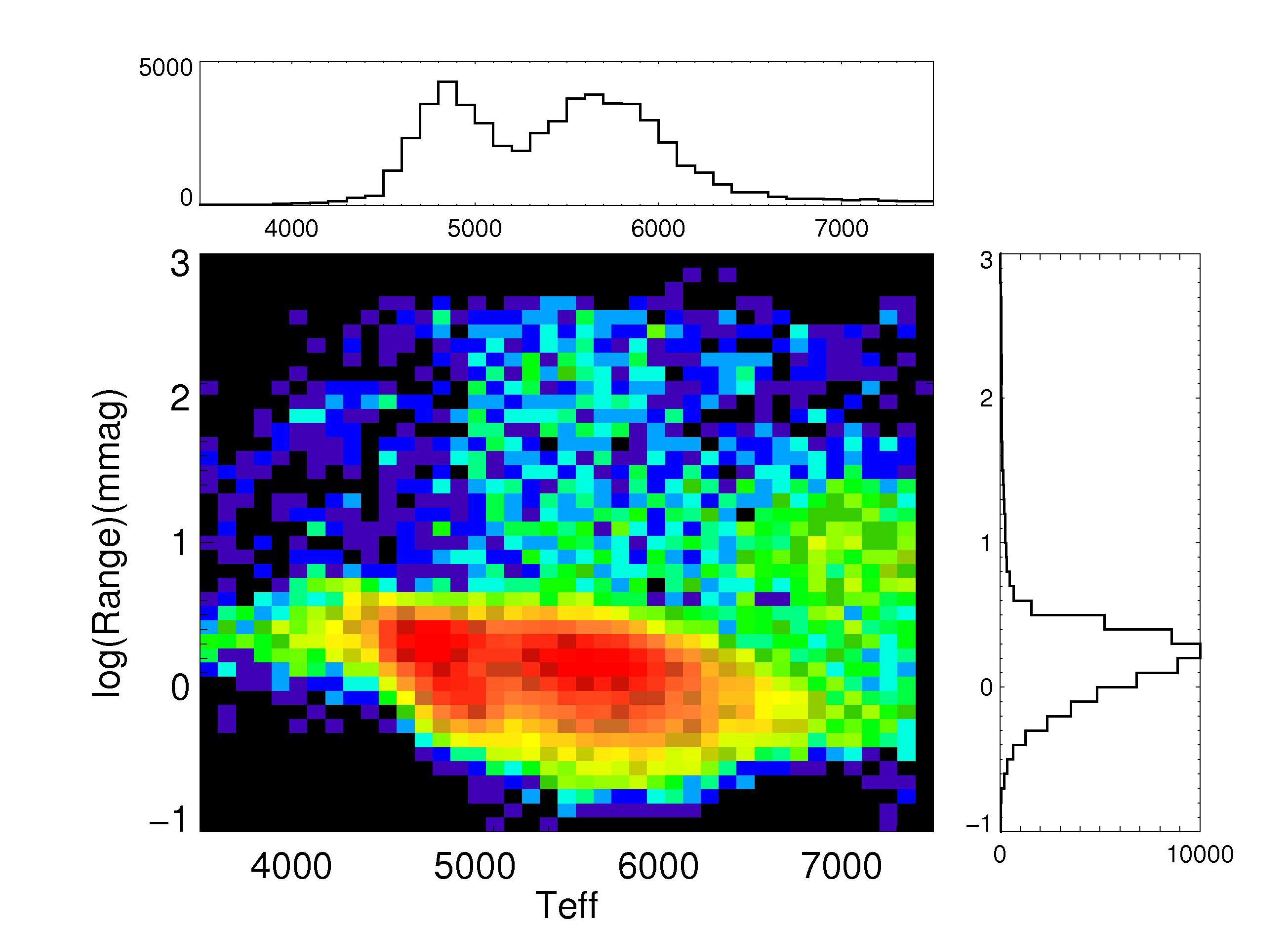}
\caption{Temperature vs. Amplitude of Variability (Range). As in Fig. 1, the upper panel shows periodic stars and the lower one non-periodic stars. The periodic stars have a cloud of higher amplitude objects at all temperatures, although there are many dwarfs with lower amplitudes for which as many are non-periodic. One can also see a tendency for cooler periodic stars to have greater amplitudes.} 
\label{f2tempvarange} 
\end{figure}

Figure \ref{f2tempvarange} shows V$_{rng}$ in the two samples as a function of effective temperature. For comparison,  V$_{rng}$ for the active Sun is about 1 mmag. The periodic sample has a cloud of points of high variability at all temperatures, and the non-periodic sample shows a two concentrations with lower variability. One is at temperatures corresponding to red giants (between 4500K$<$T$_{eff}<$5200K) and the other corresponds to less variable solar-type stars. Most of the higher amplitude objects in the non-periodic sample are hot pulsators with periods less than a day, which are not flagged by us as periodic because we have filtered for longer periods to focus on stellar rotation.

\subsection{Lightcurves of Periodic and Non-periodic Variables} 

\begin{figure} 
\centering
\includegraphics[width=0.75\textwidth]{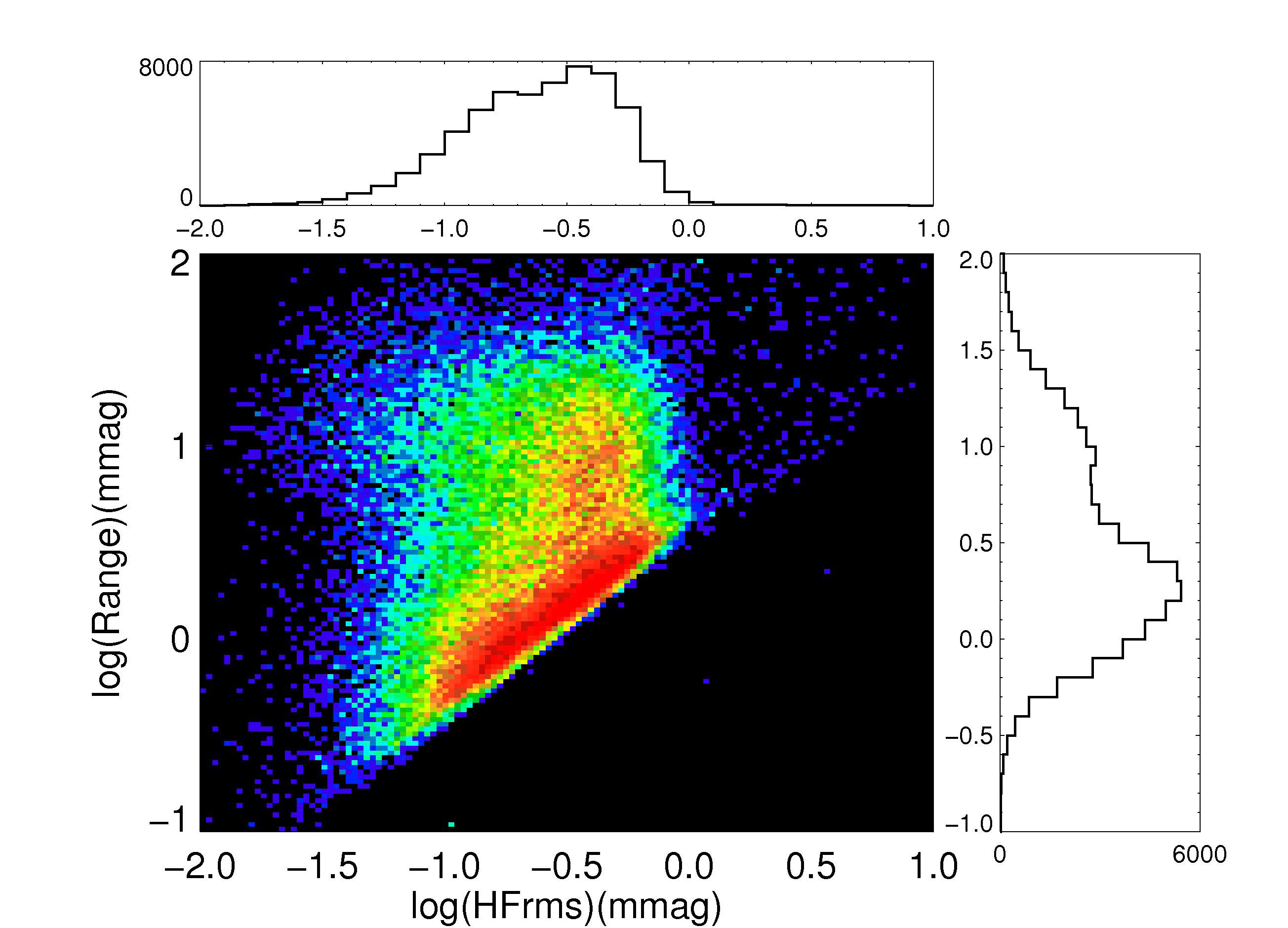} 
\includegraphics[width=0.75\textwidth]{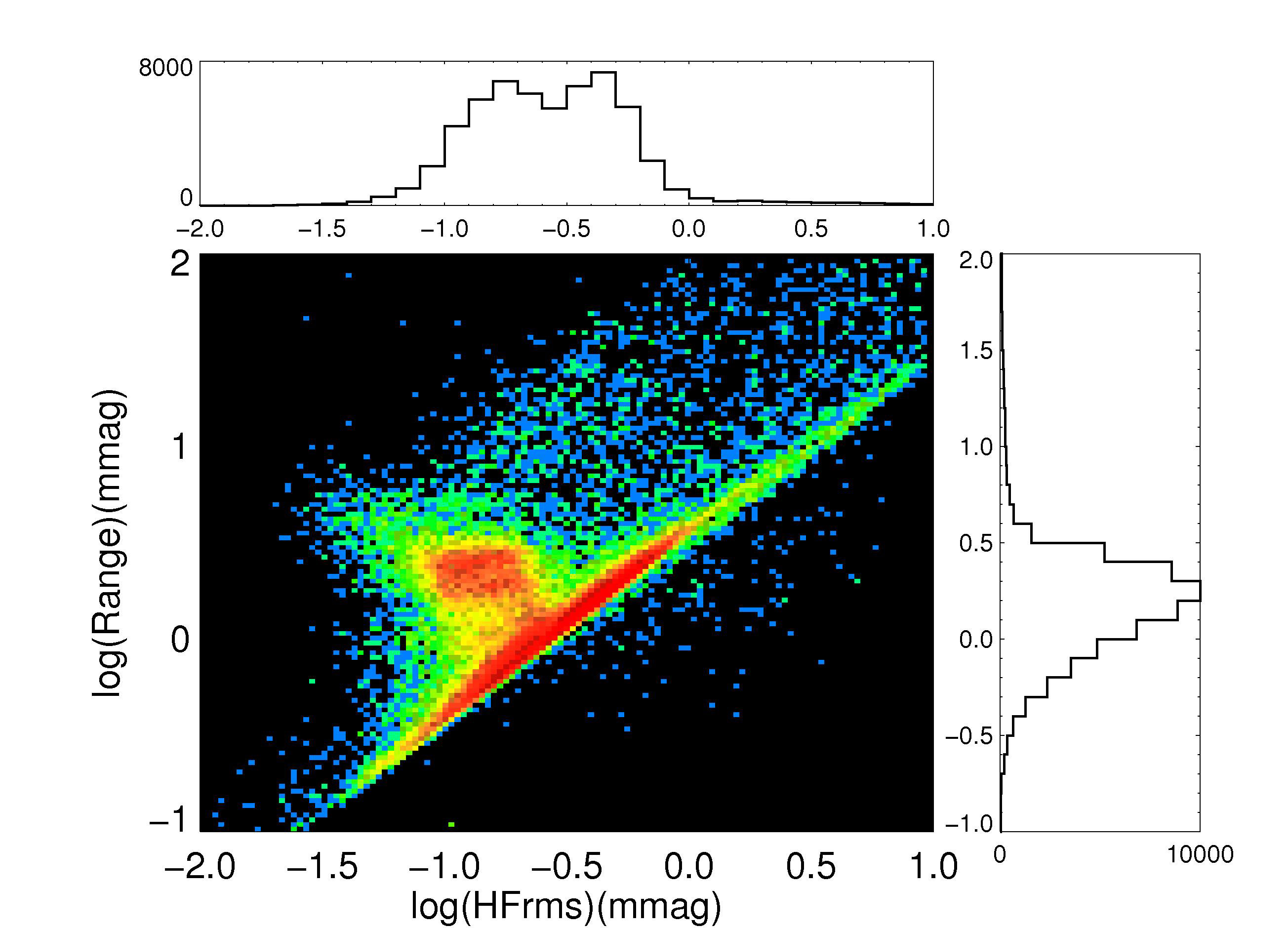}
\caption{High frequency noise vs.  Variability Range. The abscissa is a measure of the photometric noise in the target (greater noise implies a fainter target). The lower diagonal ridge is populated by stars which do not have clear variability above the noise for both the upper (periodic) and lower (non-periodic) samples. Many more periodic stars show clear variability. The rectangular feature at log(Range)$\sim$0.4 in the non-periodic sample is the locus of post-main sequence stars.} 
\label{f3rmsrange} 
\end{figure}

Another way to look at the two samples is to examine the range of variability V$_{rng}$ against the high frequency noise HF$_{rms}$ in the lightcurve due to faintness of the stars. Figure \ref{f3rmsrange} shows the comparison between them. Because of the way they are defined, V$_{rng}$ tends to be at least 4 times HF$_{rms}$ in almost all cases. This is the cause of the diagonal line along which many of the stars lie in both samples. These low-lying stars are ones for which the amplitude of variability is not much greater than the noise, although in the periodic sample a low-amplitude periodicity has been detected. Conversely, the stars lying well above the line show high-amplitude variability, whether it is periodic or not. Many of the giants show up rather clearly in the non-periodic sample at log(V$_{rng}$) between 0.2 to 0.5, and HF$_{rms}$ between -1.2 and -0.7. As a reminder, in the units of  V$_{rng}$ the depth of an earth-sized transit is about  -1.1 in the log. The objects which show up below the main diagonal are almost all cases where there are a few deep transit/eclipse dips. They don't have enough points to generate high periodogram power, and are essentially ignored by the range calculation (which drops the 5\% extremes at each end), but have an influence on HF$_{rms}$ (pushing it to the right off the main line). The periodic sample shows a much more robust cloud of higher amplitudes, at all values of HF$_{rms}$. Particularly for points above the main diagonal ridge, the lightcurves show obvious periodic variability, and in the upper ranges they almost all appear to be starspots rotating in and out of view (with a few eclipsing systems with ellipsoidal lightcurves also present).  

\begin{figure} 
\centering
\begin{center} 
\includegraphics[width=0.75\textwidth]{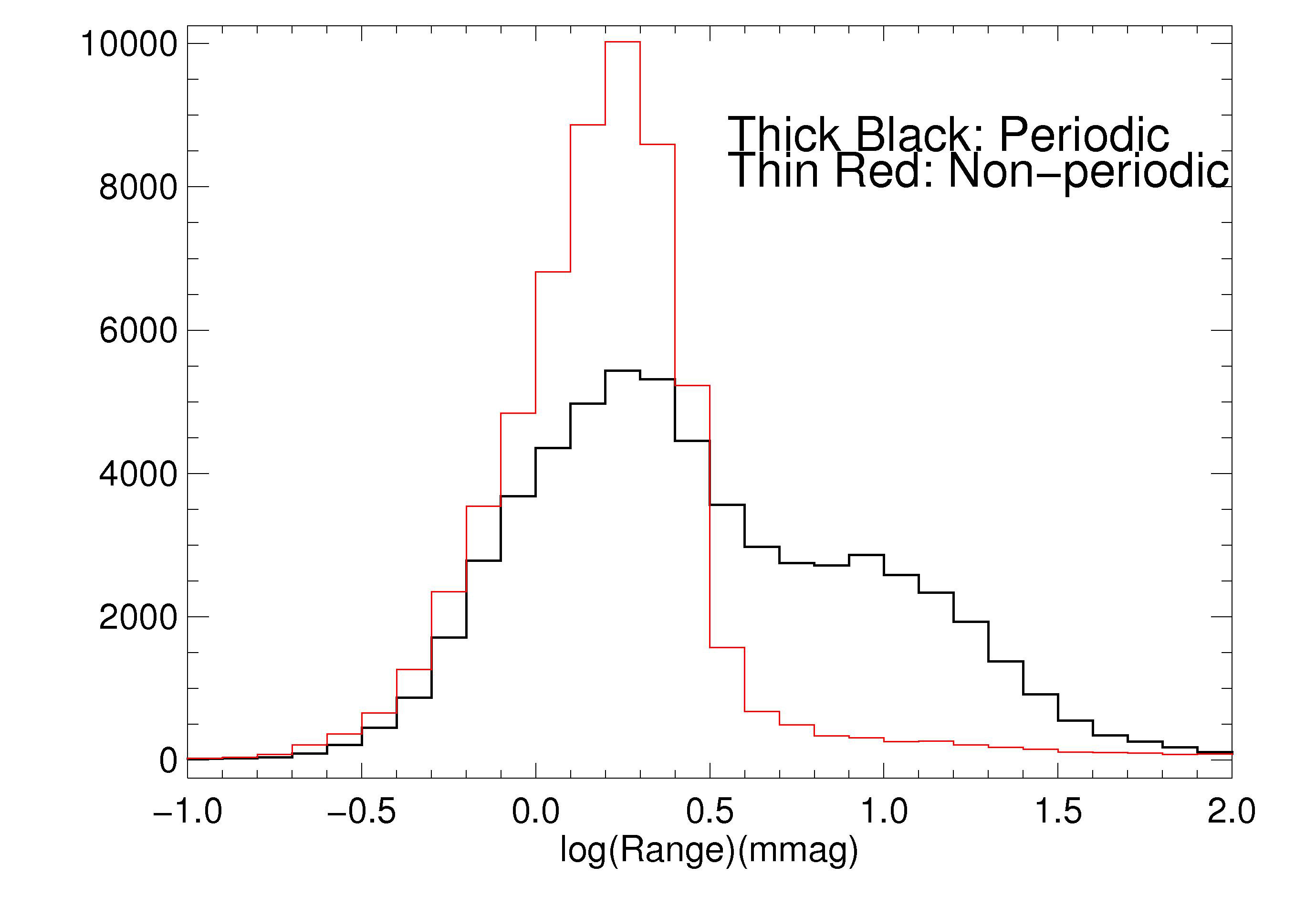} 
\end{center} 
\caption{Histograms of the amplitude of variability. The thick (black) curve is for the periodic sample of main sequence stars, and the thin (red) curve is for the non-periodic main sequence sample. This is a quantitative way to see the excess of variable stars in the periodic sample. There are similar numbers of stars with low variability in each sample (somewhat more are non-periodic). } 
\label{f4rangehist} 
\end{figure}

In Figure \ref{f4rangehist} we show the distribution of the ranges for the periodic (thicker line) and non-periodic samples. This conveys some of the same information as in Fig. \ref{f3rmsrange}, but shows more clearly that the periodic sample has many more stars with high-amplitude variability (as would be expected if starspots were the source of the variability). The distributions are much more similar at the low-amplitude end. Here the stars are quiet (don't show much variability); a little more than half of those are non-periodic. We have used spot modeling to convert an amplitude into a spot covering fraction, assuming spot contrast of about 0.6 (which may not hold over all spectral classes). This yields the result that  log( V$_{rng}$)=0.3 corresponds to a spot coverage of 1\% of the stellar surface. A simple estimation using half-black spots gives about the same answer. The active solar coverage is about 0.5\%, the peak of the distribution here is at about 1\%, and more than half of the stars flagged as periodic have coverages well above 1\%, extending up to 20\% or more in the largest cases. Some of those extreme cases are not spots but ellipsoidal binaries; others are indeed very spotted stars (judging from the variations in the lightcurve from cycle to cycle, which binaries should not exhibit as readily). 

\begin{figure}
\centering 
\includegraphics[width=0.49\textwidth]{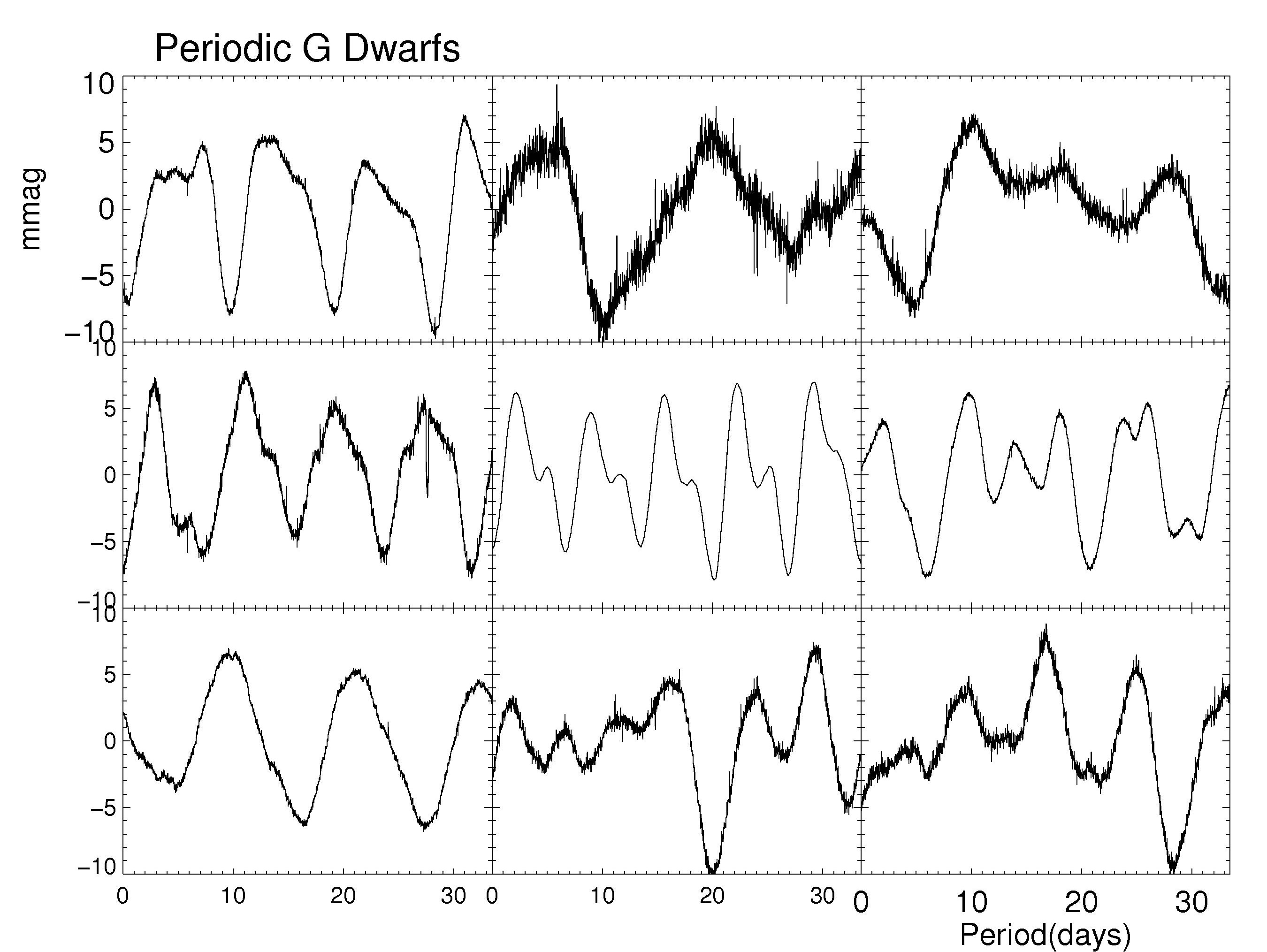} 
\label{f5PGdwarf} 
\includegraphics[width=0.49\textwidth]{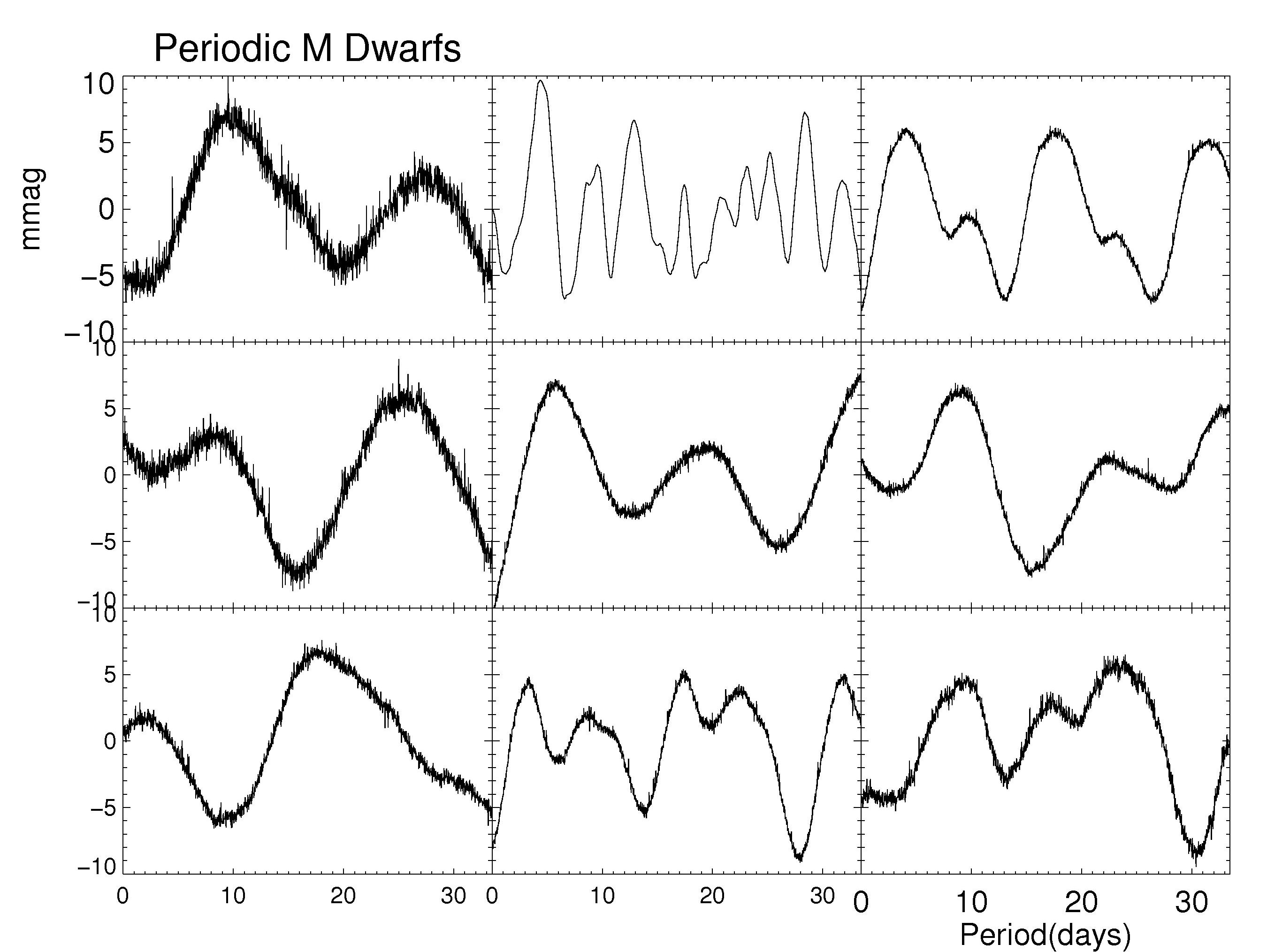} \\
\label{f5PMdwarf} 
\includegraphics[width=0.49\textwidth]{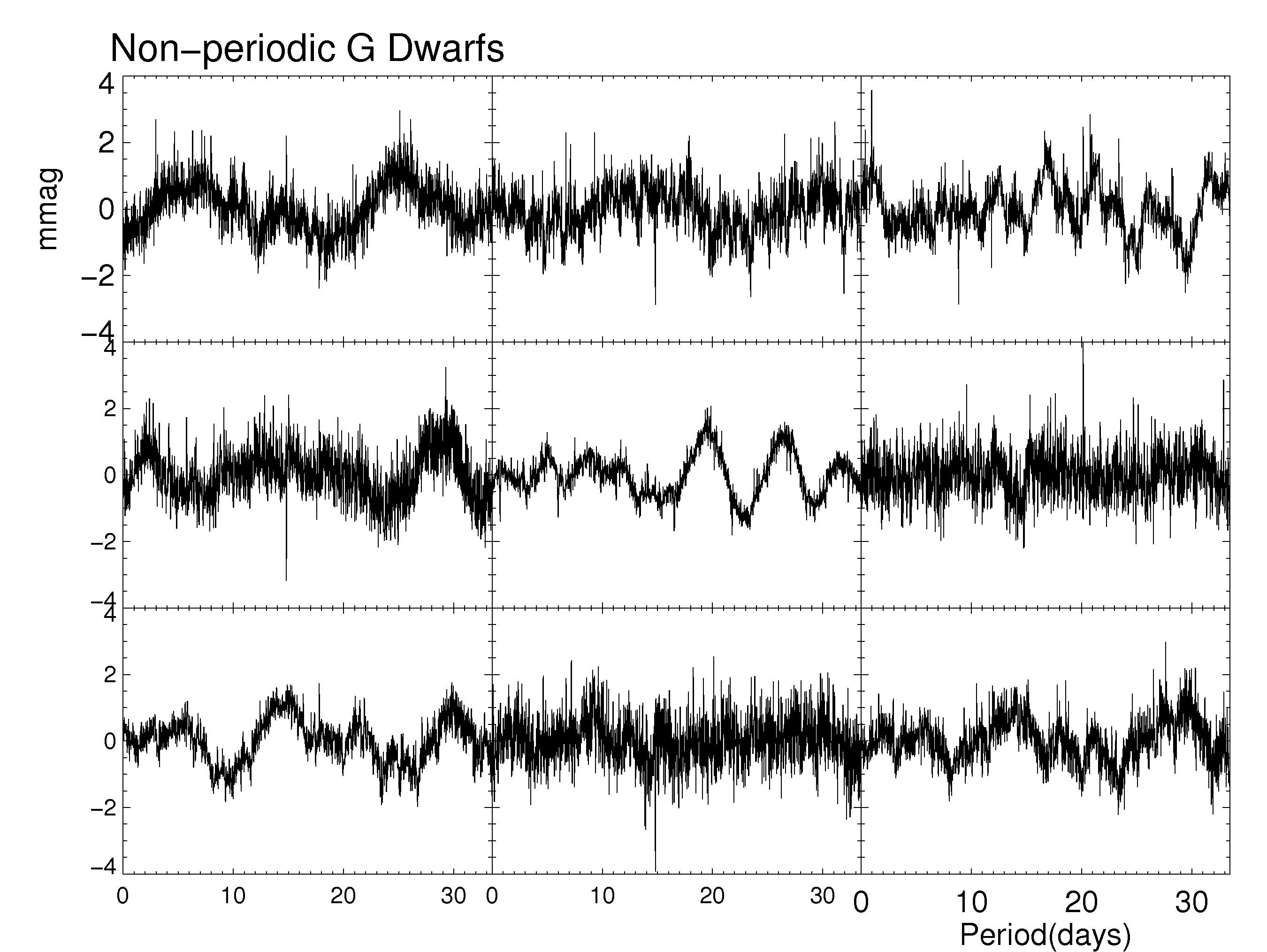} 
\label{f5nonPGdwarf} 
\includegraphics[width=0.49\textwidth]{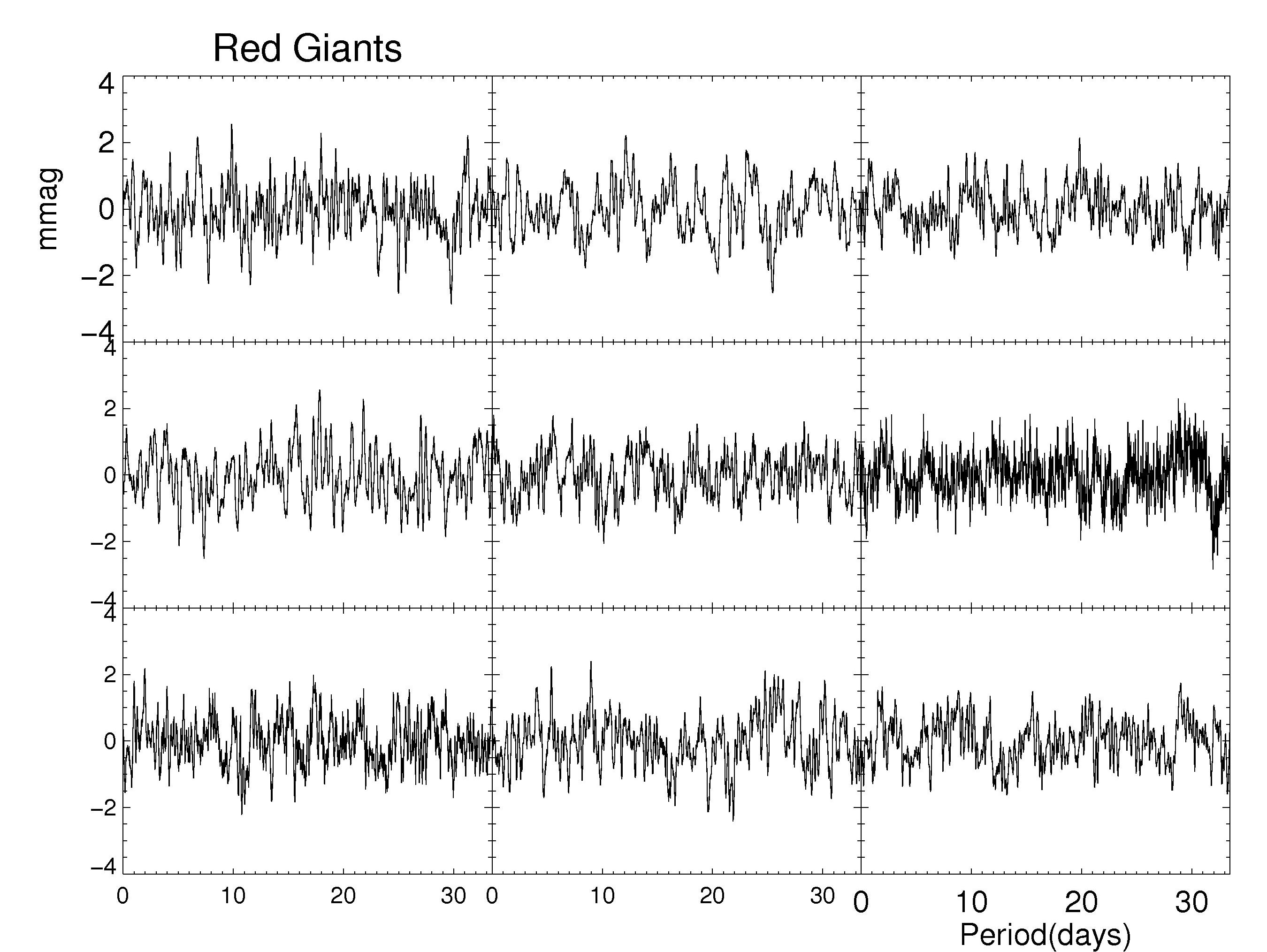} 
\caption{  Upper left: Lightcurves for clearly periodic G dwarfs (see text for a more exact description of stellar selection criteria).  Kepler IDs for these stars are:  5302013, 9401951, 8367679, 11773022, 7200111, 4640983, 7037146, 12069336, 9269023. Upper right: Lightcurves for clearly periodic M dwarfs.  Kepler IDs for these stars are: 11775907, 2424191, 4743351, 6382217, 6420895, 8558589, 9573685,5950024, 4554367. Lower left: Lightcurves for weakly periodic G dwarfs. Note change of scale. Kepler IDs for these stars are: 9580212, 6125701, 8308260,  6837899, 10858832, 4366093, 4552939, 11598724, 3110216. Lower right: Lightcurves for red giants. Kepler IDs for these stars are: 12204548, 3427850, 10666932, 12208273, 4772722, 4725874, 10937855, 9783225, 4271855. } 
\label{f5Giants} 
\end{figure}

We now present some typical lightcurves. In Fig. \ref{f5Giants}a are some obviously periodic G dwarfs. The parameter space from which they are randomly chosen is 5800K$<$T$_{\rm eff}<$5850, log(g)$>$4.2, 8$<$V$_{rng}<$12,100$<$ P$_{str}<$120, and period (days) between 4 and 12. Fig. \ref{f5Giants}b shows a set of M dwarfs with the same parameters (except 3900K$<$T$_{\rm eff}<$4000K). One can see from the noise which stars are fainter (the top middle M dwarf is substantially brighter than the others). There is not an obvious difference between the G and M stars, but of course we've chosen them from the same relevant parameter spaces (although it is also true that the G and M dwarfs don't populate these spaces very differently). In Fig. \ref{f5Giants}c we return to the G dwarfs, but now look in the region 2$<$V$_{rng}<$4 and 65$<$P$_{str}<$70. The intensity scale has been decreased by a factor of 2 to better display the decreased range. In most cases the periodicity is still obvious. In a couple of instances it is harder to pick out a periodicity by eye. Finally, we present some lightcurves of stars in the giant domain in Fig. \ref{f5Giants}d. These have  4700K$<$T$_{\rm eff}<$4750K, 2.5$<$log(g)$<$3, 2$<$V$_{rng}<$4, and 20$<$P$_{str}<$25. They look different from the G dwarfs in that the variability is aperiodic and fairly consistent in amplitude. 

\begin{figure} 
\centering
\includegraphics[width=0.75\textwidth]{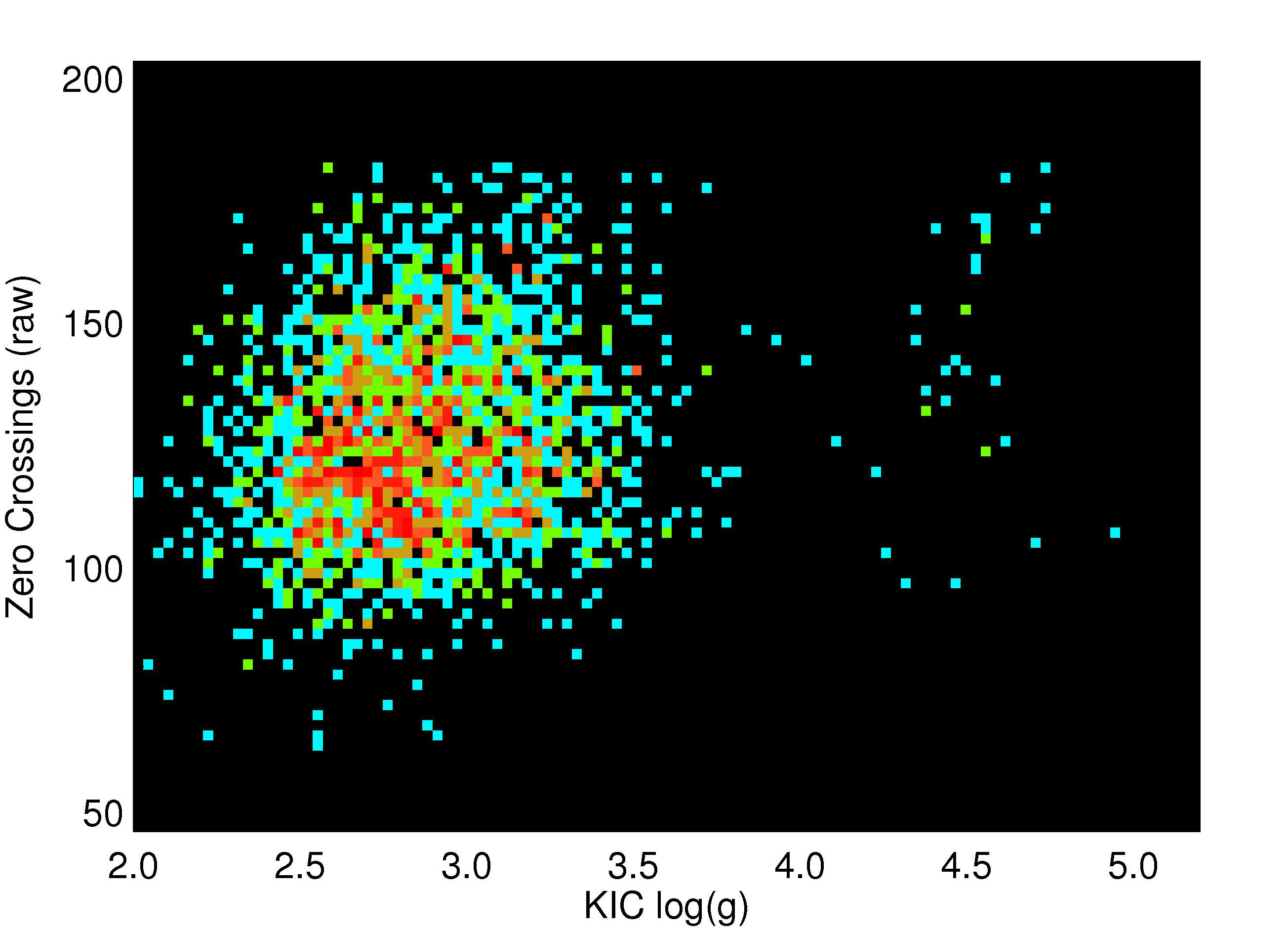} 
\caption{The results of a photometric filter (see text) intended to select giants but not dwarfs. The ordinate is one filter that was employed (the number of times the differential lightcurve crossed zero. The KIC gravities were not used in choosing the stars, but afterwards to see whether giants were chosen. The high gravity cases look very much like giants photometrically (and it is possible that some were misclassified by the KIC). } 
\label{f6photgiant} 
\end{figure}

We investigated whether one can isolate giants purely by their photometric behavior. It has been well established that they have a characteristic photometric signature \cite{GillGiant}; the question we are asking here is how well one can eliminate dwarfs which might also show similar behavior (even though the vast majority of them do not). We provide one reasonably successful attempt to do this without using KIC gravities as priors for G,K,M stars (we use them to check afterward). We choose stars with V$_{rng}$ between 1 and 5 (motivated by where they are seen in Fig. \ref{f3rmsrange}), and with the number of peaks within 10\% of the primary peak power greater than 10 (based on the periodograms of known giants). Finally we pick the number of smoothed zero crossings to be greater than 40 (to pick out the right timescale for giant photometric excursions) but the number of unsmoothed zero crossings to be less than 200 (to eliminate noise as the primary source of zero crossings). Figure \ref{f6photgiant} shows the result (with the unsmoothed zero crossings as the ordinate and KIC gravity as the abscissa). Most of the giants pass through our filter while very few dwarfs do. The remaining dwarfs have lightcurves that indeed closely resemble giant lightcurves, and perhaps some of them are actually misclassified giants. This can only be resolved through ground-based spectroscopy, or if the promise of detecting all the dwarf parallaxes from \textit{Kepler} data itself is realized. 

\subsection{Periodic Spotted Main-sequence Stars} 

\begin{figure} 
\centering
\includegraphics[width=0.75\textwidth]{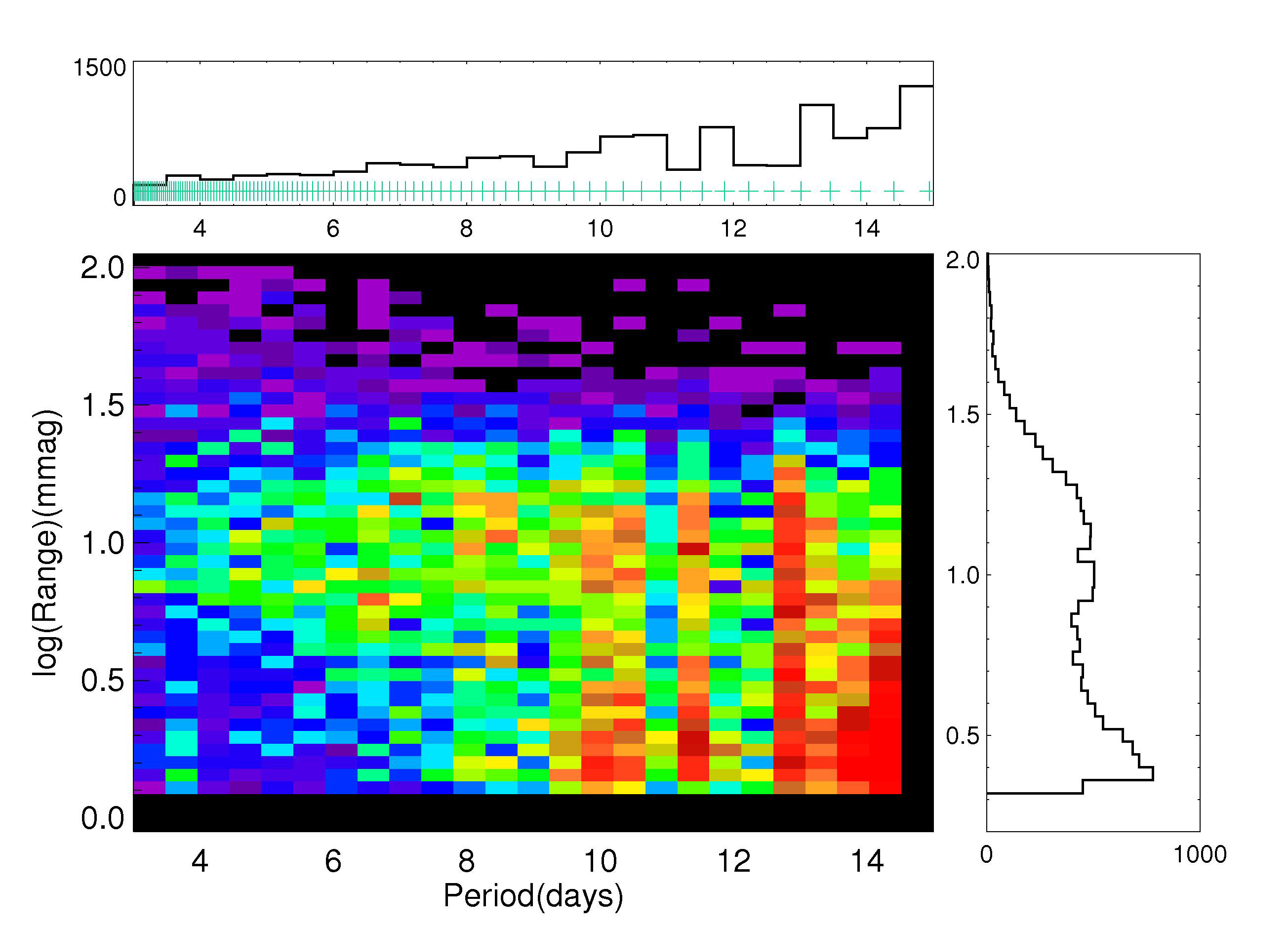} 
\caption{The distribution of periods found for a selected sample of periodic main sequence stars (see text). The actual sampling of the periods search is given by the row of plusses in the upper histogram. The period histogram rises toward our limit of validity of 16 days. The distribution of variability range peaks just above the active Sun value, with a substantial tail at high variability (which extends to shorter periods). } 
\label{f7perpowhist} 
\end{figure}

We now take a look at the periodic sample in more detail. Figure \ref{f7perpowhist} shows the distribution of primary periods within a sample that has been further restricted by choosing only stars with KIC gravities greater than log(g)=4 (which still leaves over 50,000 stars in the sample). The distribution of periods rises towards 2 weeks. Beyond that we do not trust the results because the time coverage is only 34 days, and because removing a third order fit can definitely have an influence on whether the lightcurve shows features timescales longer than 2 weeks. The small peak at about 3.5 days might possibly be due to the fact that one of the guide stars was a spotted star with this period (see the Kepler Data Release 5 Notes; it was removed after Q1) which caused the pointing to vary periodically. The structure at longer periods is due primarily to the fact that the sampling of periods tested is becoming sparser (the actual sampling is given by the row of cyan plusses in the upper histogram). We also investigated whether the range of variability is a function of the primary period. Generally, shorter periods indicate more active stars \cite{Pizzolato}, and these might be both more obviously periodic or show larger amplitudes of variation (unless the activity were too uniformly distributed). We see that the bulk of the stars have  V$_{rng}$ a little greater than the active Sun, and there is modest evidence of greater variability at shorter periods (the lowest V$_{rng}$ are found mostly at the long periods) . Of course, by restricting this study to stars rotating fast enough to show periodicity in two weeks, we have biased the result against the slower rotators (many of which presumably currently lie in our non-periodic sample). It is known that activity saturates at short periods and that might be playing a role here. Still, it is perhaps a little surprising that there is not a more obvious effect, especially since we are sampling rotation periods up to solar in the case of a half-period of two weeks. 

We remind the reader that we did not analyze periods below 2 days; there is indeed a population of short period cases but we expect that  those are not due to rotation (except perhaps in close binaries). It is important to note that one cannot assume that the dominant period is the rotation period of the star, even when the signal looks very much like starspots. In many cases it is more likely to be the half-period; when a star has an asymmetric spot distribution each face produces a dip of a different depth so that there could be two major dips per rotation. If there is essentially one active location, or a continuous enough distribution around the star, two discrete dips may not appear and then the periodogram will find the actual period.

To produce an active Sun sample for comparison, 20 segments of SOHO DIARAD \cite{SOHO} data from 2001 were recast to have similar cadence and length as the \textit{Kepler} data (as in Paper I). Periods are found between 10-50 days with a mean and median of about 20 days (and mean P$_{str}$ of ~200). Of course, most of these periods (even those matching the actual solar period) would not be considered valid in our current analysis (since they are derived from 33 day data segments and we don't trust periods over 16 days). We composed a similar sample of quiet Sun data from 1996 SOHO  measurements, and actually obtained a similar period distribution with a mean P$_{str}$ of about 150. The high significance of P$_{str}$ is primarily due to the very low noise in the SOHO data. We performed the additional experiment of removing a third order fit from the quiet Sun data; this reduced P$_{str}$ to a little below 100 but left the dominant period distribution as rather similar.  

\begin{figure} 
\centering
\includegraphics[width=0.75\textwidth]{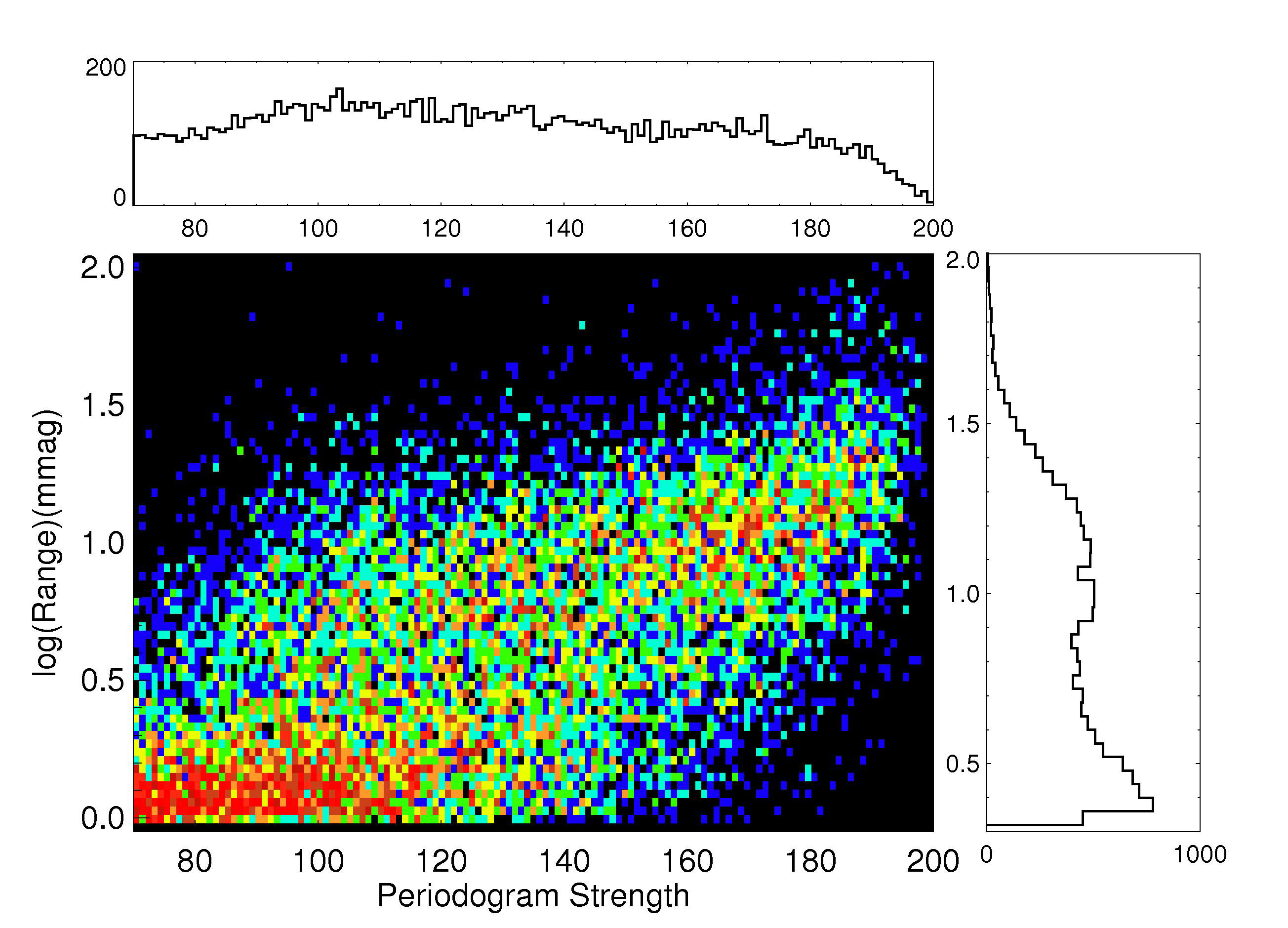} 
\label{f8dwfrotators} 
\includegraphics[width=0.75\textwidth]{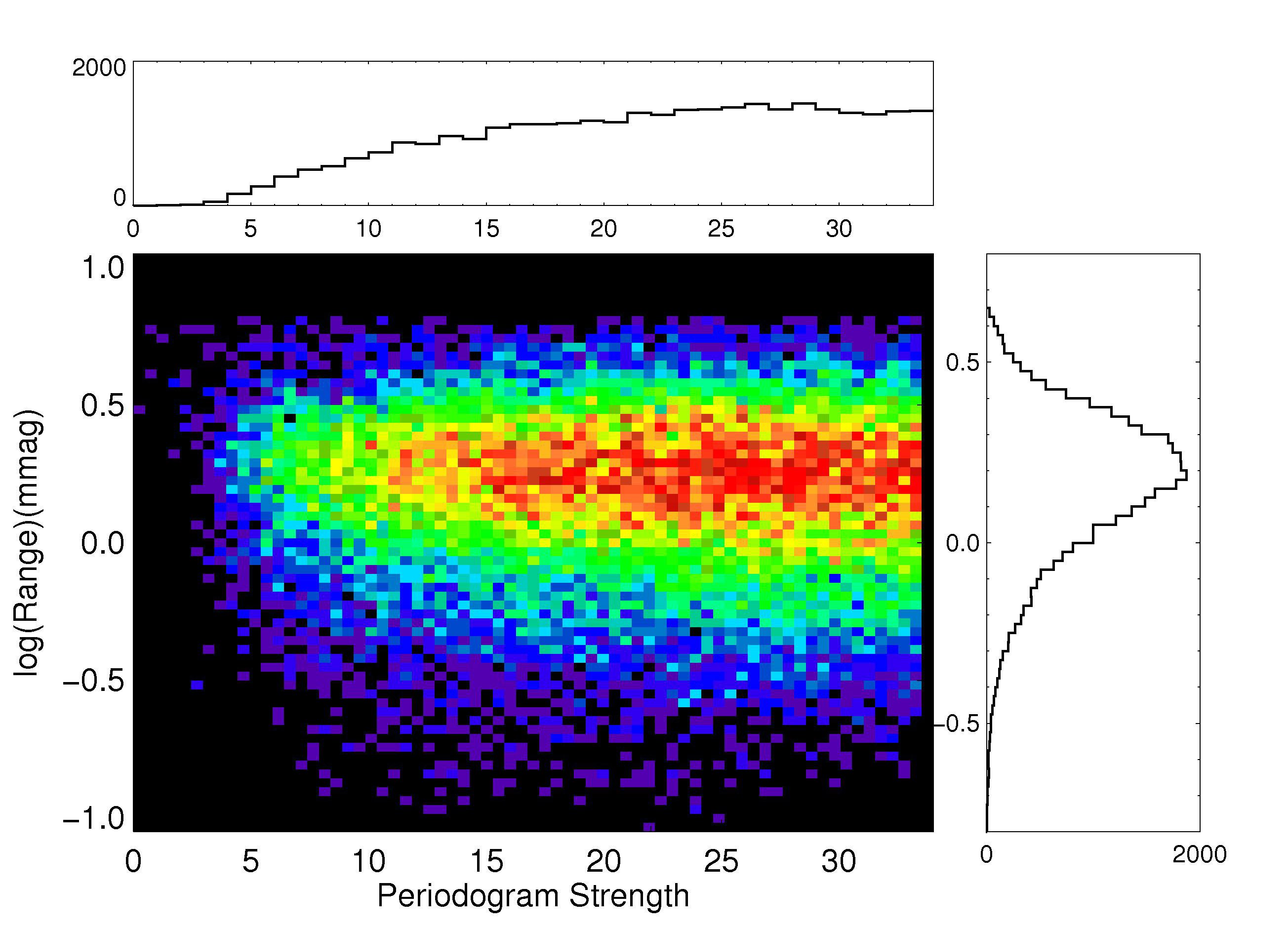} 
\caption{Periodogram strength vs. variability amplitude in the periodic sample (upper panel). There is a clear tendancy for these to be correlated, which is not surprising, but it does indicate that very large amplitude variations tend to be more strongly periodic. There is also quite a spread of amplitudes at a given value of periodogram strength. The lower panel shows the non-periodic sample. Note the change in scale for the abcissa. The amount of variability tends to be low, and there is no relation between it and what periodogram power is present (formally, values above 15 have false alarm probabilities $<10^{-5}$).  }
\label{f9nonpwrrng} 
\end{figure}

 In order to better study this, we selected a subsample out of the periodic sample by restricting it to KIC gravities greater than log(g) of 4.2, temperatures between 3800K and 6200K, periods between 3 and 16 days, P$_{str}>$70, and V$_{rng}>$2. The upper panel of Figure \ref{f9nonpwrrng} shows some trend in this sample of these nearly 14,000 clear dwarf rotators for P$_{str}$ to increase with V$_{rng}$, but there is a great deal of scatter along both axes.   In this periodic dwarf sample there is a general decline in population to higher periodogram strengths, but all were chosen to show very clear periods. The analogous plot for the non-periodic sample (lower panel) is quite different, showing again that periodic stars tend to be more variable (although there is a region of low variability in common). Even these non-periodic stars lie primarily above the formal criterion for a significant period (which is at  P$_{str}\sim$15). For solar-type main sequence stars, high variability very likely arises from starspots (which then exhibit periodicity through stellar rotation), and objects which don't exhibit periods (or half-periods) as short as 2 weeks are not as variable. We often see possible signs of differential rotation in the variable stars (even with such a short span of data). Such signs are typically that there are two dips of different depth, and they can be seen to move out of phase with respect to each other. An excellent longer span example of this is shown by CoRoT-Exo-2b \cite{Exo2b}. The dominant period can change as the spot distribution changes (although we don't have long enough data segments here to really see that).

\section{Conclusions and Future Work} 

We find that the amplitude of variability is generally substantially larger for stars that are clearly periodic than for those that are not. Among the non-periodic variables, giants are the dominant constituent. We are able to select for giants based on their photometric properties alone with reasonable efficiency. The high amplitude non-periodic stars are almost all high frequency pulsators (too high for our tailored periodogram test to pick them up). For the periodic variables, the dominant period distribution rises to the long end of what we consider well-determined (up to two weeks). Since a substantial fraction of stars with dominant periods of two weeks may actually have rotation periods of a month, we are sampling into the pool of main sequence stars with solar rotation. Some of the stars with even longer periods must wait for more time coverage, and some of them are quiet enough to have been classified as non-periodic in the current study. It is too early to give a real rotational period distribution or make statements about the age distribution in this sample. 

Perhaps the main conclusion of this preliminary overview of stellar variability as seen by the  \textit{Kepler} mission is that this will be a very rich and unique dataset for studying the surfaces of stars. This study is preliminary both because the data reduction will require further refinement and mostly because we have only treated the first month of what is already more than a year (expected to become at least 3.5 years) of nearly continuous measurements at very high precision of roughly 150,000 stars. Well over half of the sample of main sequence solar-type stars show variability at a level beyond the noise, and many of them show apparently periodic behavior (60,000 more clearly and 34,000 somewhat marginally). Even restricting the sample to solar-type main sequence stars with strong amplitudes and periodicities produces a population of 14,000.  

We have only looked by eye at a small minority of them so far. The stars whose variability is periodic but clearly not due to spots (pulsators and eclipsers are the primary other variables) are both far less numerous and can be filtered out (or selected) with good efficiency. While it is quite obvious in many cases that we are looking at starspots rotating in and out of view, evolving in strength, and differentially rotating at different latitudes, more work is needed to distinguish between these and other sources of variability in more ambiguous cases. We believe that the ambiguity of some of the lightcurves will be reduced or eliminated as longer time series become available. A lot of additional work is needed to model the spotted cases sufficiently well to extract all the information that is present. This will allow a qualitative leap in our understanding of magnetic activity on stars.

\acknowledgements 
The authors wish to thank the entire Kepler mission team, including the engineers and managers who were so pivotal in the ultimate success of the mission. LW is grateful for the support of the \textit{Kepler Fellowship for the Study of Planet-Bearing Stars}. GB thanks the NSF through grant AST-0606748 for partial support of this work. Funding for this Discovery mission is provided by NASA's Science Mission Directorate. 

{\it Facilities:} The \textit{Kepler} Mission

\end{document}